\pdfoutput=1

\documentclass[11pt]{article}

\usepackage[preprint]{acl}

\usepackage{times}
\usepackage{latexsym}

\usepackage[T1]{fontenc}

\usepackage[utf8]{inputenc}

\usepackage{microtype}

\usepackage{inconsolata}

\usepackage{graphicx}
\usepackage{longtable}

\usepackage{color,soul}
\usepackage{tabularx}
\usepackage{booktabs} 
\usepackage{natbib}
\usepackage{multirow} 
\usepackage{array}
\usepackage{makecell}
\usepackage{array}
\usepackage{geometry}
\usepackage{xcolor}

\definecolor{cyellow}{HTML}{FCEBB9}
\definecolor{cpurple}{HTML}{C4D0FF}
\definecolor{cpink}{HTML}{EADBE7}
\definecolor{cblue}{HTML}{D3EEF9}

\title{Towards a Design Guideline for RPA Evaluation: \\ A Survey of Large Language Model-Based Role-Playing Agents}

\author{Chaoran Chen$^{\dagger}$ \\
\small{University of Notre Dame}
\And Bingsheng Yao$^{\dagger}$ \\
\small{Northeastern University}
\And Ruishi Zou \\
\small{University of California, San Diego}
\AND Wenyue Hua \\
\small{University of California, Santa Barbara}
\And Weimin Lyu \\
\small{Stony Brook University}
\And Yanfang Ye \\
\small{University of Notre Dame}
\AND Toby Jia-Jun Li \\
\small{University of Notre Dame}
\And Dakuo Wang 
\thanks{Corresponding author: d.wang@northeastern.edu \\ 
\indent ~$^{\dagger}$ Equal contribution. \\ 
\indent ~~Github repository: \url{https://github.com/CRChenND/LLM_roleplay_agent_eval_survey} \\
\indent ~~Searchable webpage: \url{https://agentsurvey.hailab.io/}
}\\
\small{Northeastern University}
}

\begin{document}
\maketitle

\begin{abstract}
Role-Playing Agent (RPA) is an increasingly popular type of LLM Agent that simulates human-like behaviors in a variety of tasks. 
However, evaluating RPAs is challenging due to diverse task requirements and agent designs.
This paper proposes an evidence-based, actionable, and generalizable evaluation design guideline for LLM-based RPA by systematically reviewing $1,676$ papers published between Jan. 2021 and Dec. 2024.
Our analysis identifies six agent attributes, seven task attributes, and seven evaluation metrics from existing literature.
Based on these findings, we present an RPA evaluation design guideline to help researchers develop more systematic and consistent evaluation methods.

\end{abstract}


\begin{figure}[t]
    \centering
    \includegraphics[width=.98\linewidth]{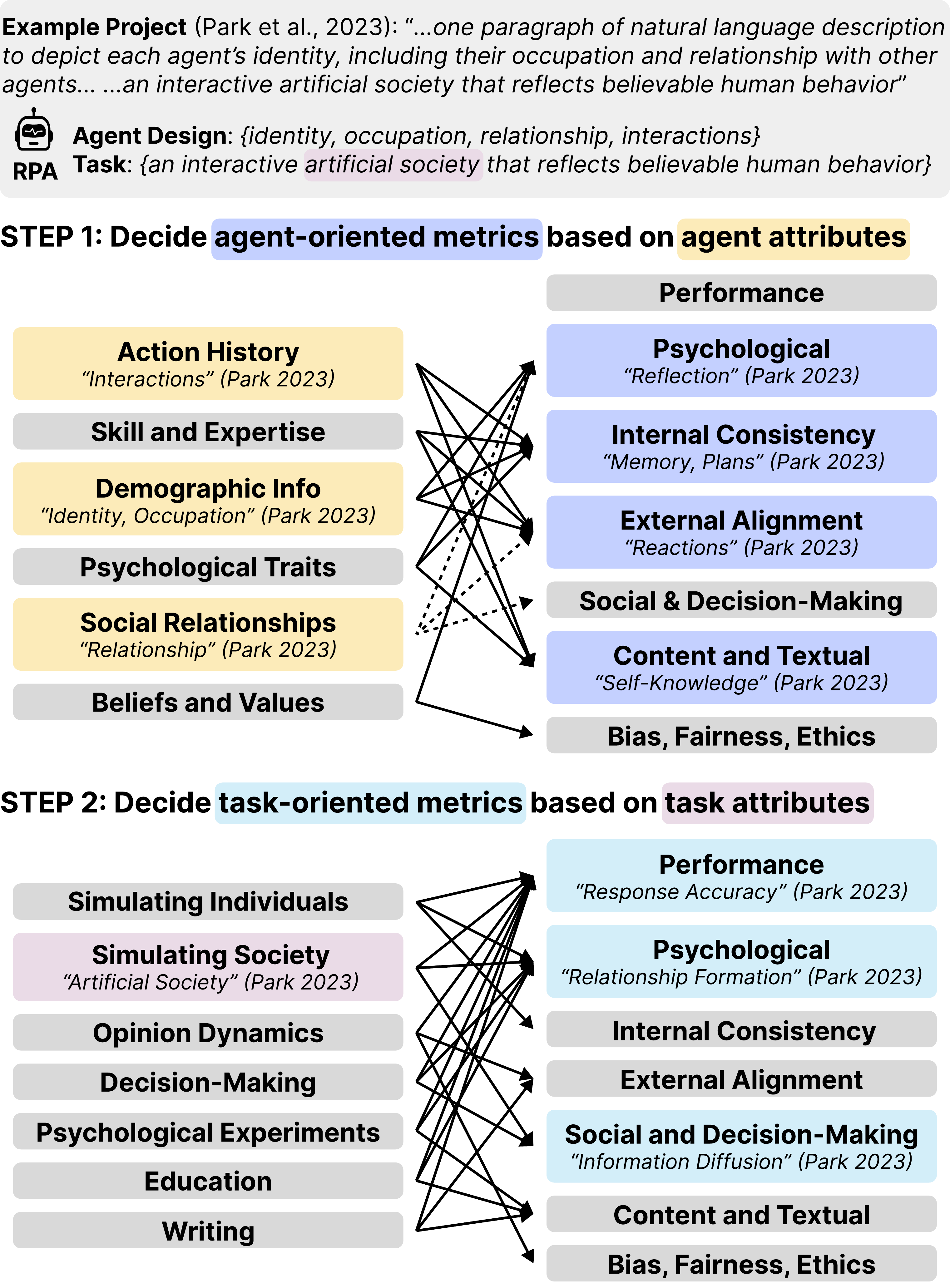}
    \vspace{-0.5em}
    \caption{RPA evaluation design guideline. To illustrate how to use it in practice, we pretended we were selecting the evaluation metrics for the  ``Stanford Agent Village''~\cite{park2023generative} given agent attributes (yellow) and task attributes (pink). The original authors' selection of evaluation metrics (purple and blue) perfectly aligns with our RPA design guideline, which echoes their work's robustness. More details in Sec~\ref{sec_good_example} and a bad example in Sec~\ref{sec_bad_example}.}
    \vspace{-1.5em}
    \label{fig:teaser}
\end{figure}

\section{Introduction}
LLMs have yielded human-like performance in various cognitive tasks (e.g., memorization~\cite{schwarzschild2025rethinking}, reasoning~\cite{wang2023can, plaat2024reasoning}, and planning~\cite{song2023llm, huang2024understanding}). 
These emergent capabilities have fueled growing research interest on 
\textbf{Role-Playing Agent} (RPA)~\cite{chen2024from,tseng-etal-2024-two}:
RPAs are digital intelligent agent systems powered by LLMs, where users provide human-like \textbf{agent attributes} (e.g., personas) and \textbf{task attributes} (e.g., task descriptions) as input, and prompt the LLM to generate human-like behaviors and the reasoning process. 
The potential of RPAs is promising and far-reaching, as illustrated by the early results of the massive interdisciplinary studies in social science~\cite{10.1145/3526113.3545616, park2023generative}, network science~\cite{10.1145/3613904.3642363}, psychology\cite{jiang-etal-2024-personallm} and juridical science~\cite{he-etal-2024-agentscourt}.

Despite growing interest in RPAs, a fundamental question remains: \textbf{how can we systematically and consistently evaluate an RPA?}
How should we select the evaluation metrics, so that the evaluation results can be comparable or generalizable from one task to another task?
Addressing these challenges is difficult~\cite{dai2024mmrole, tu2024charactereval, wang2024incharacter}. 
due to the vast diversity of tasks (e.g., simulating an individual's online browser behavior~\cite{10.1145/3613904.3642363} or simulating a hospital~\cite{li2024agent}), and the high flexibility in RPA design (e.g., an agent persona can be one sentence or 2-hours of interview log~\cite{park2024generativeagentsimulations1000}).
Another challenge is the inconsistent and often arbitrary selection of evaluation methods and metrics for RPAs, raising concerns about the validity and reliability of evaluation results~\cite{wang2025limits, zhang2025simulation}.
As a result, the research community finds it difficult to compare the performance across multiple RPAs in similar tasks reliably and systematically.

To address this gap, we propose an evidence-based, actionable, and generalizable design guideline for evaluating LLM-based RPAs. We conducted \textbf{a systematic literature review} of $1,676$ papers on the LLM Agent topic and identified $122$ papers describing its evaluation details. 
Through expert coding, we found that agent attribute design interacts with task characteristics (e.g., simulating an individual or simulating a society requires a diverse set of agent attributes). 
Furthermore, we synthesized common patterns in how prior research successfully (or unsuccessfully) designed their evaluation metrics to correspond to the RPA's agent attributes and task attributes. 
Building on these insights, we propose an RPA evaluation design guideline (Fig.~\ref{fig:teaser}) and illustrate its generalizability through two case studies. 
\section{Related Work}

\begin{figure}[t]
    \centering
    \includegraphics[width=.98\linewidth]{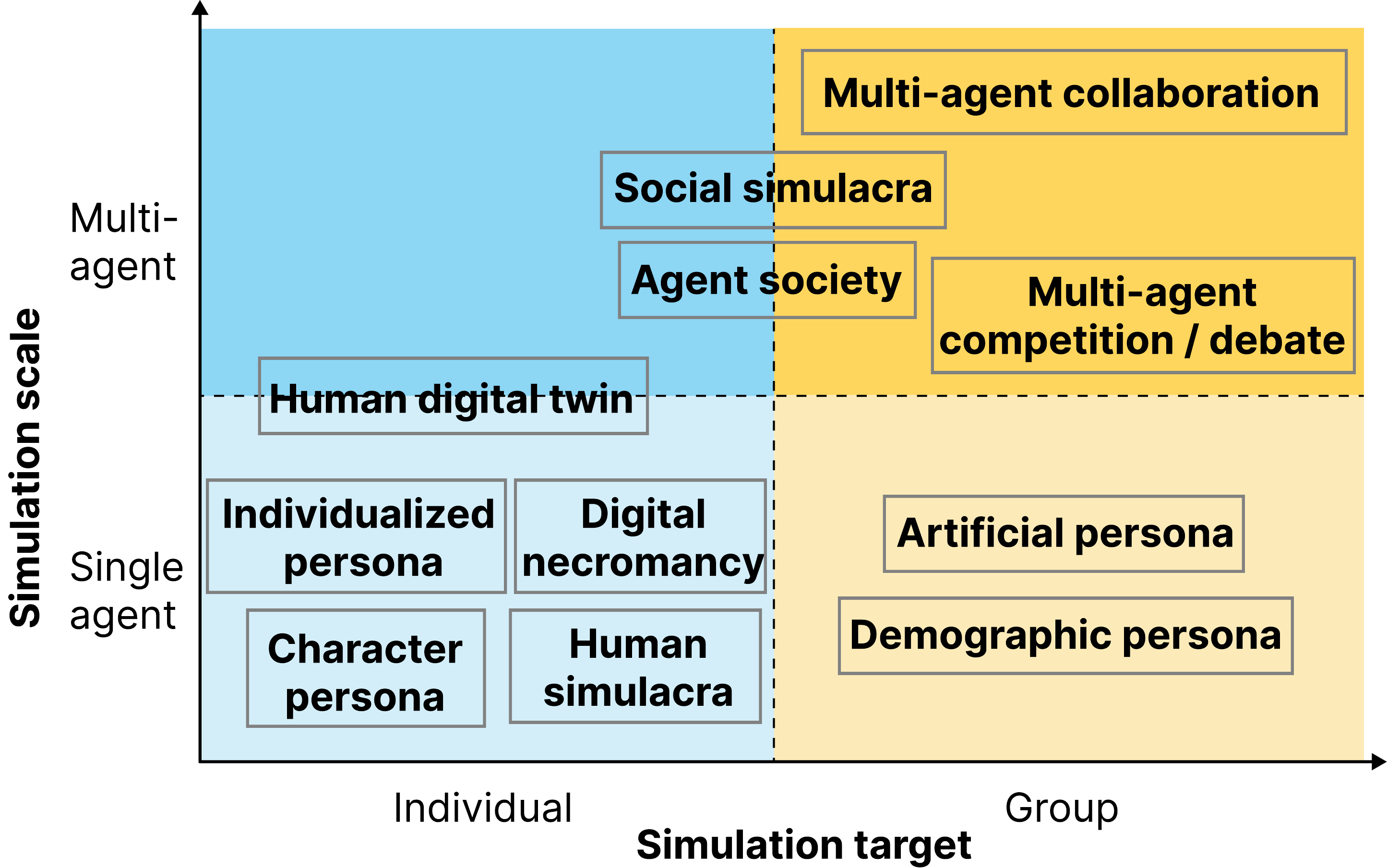}
    \vspace{-0.5em}
    \caption{Taxonomy of RPAs.}
    \vspace{-1.5em}
    \label{fig:rpa-taxonomy}
\end{figure}

\subsection{Taxonomy of RPAs}
Existing literature~\cite{chen2024from,tseng-etal-2024-two,chen2024oscars,mou2024individual} classifies RPAs along two independent dimensions: Simulation Target and Simulation Scale. The Simulation Target dimension differentiates between agents that simulate specific individuals (e.g., historical figures, fictional characters, or individualized personas) and those that simulate group characteristics (e.g., artificial personas)~\cite{chen2024from,tseng-etal-2024-two,chen2024oscars}. The Simulation Scale dimension categorizes agents by the complexity of their interactions, ranging from single-agent simulations with no social interaction to multi-agent systems that replicate structured or emergent societal behaviors~\cite{mou2024individual}.

To unify these perspectives, we introduce an integrated taxonomy for RPAs (Fig.\ref{fig:rpa-taxonomy}). The \textit{Simulation Target} axis distinguishes between individual-focused and group-focused agents. Examples of individual-focused agents include digital twins, which model an individual's decision-making process~\cite{rossetti2024social}, and personas, which emulate specific human-like characteristics~\cite{10.1145/3613904.3642363}. Group-focused agents include social simulacra, which model interactions between specific individuals within a group (e.g., the relationship dynamics in Detective Conan)~\cite{wu2024role}, and synthetic societies, which replicate large-scale social structures and emergent group behaviors~\cite{park2023generative}. The \textit{Simulation Scale} axis differentiates between single-agent and multi-agent systems. Single-agent RPAs operate at an individual level, such as digital twins used for personalized recommendations or personas that generalize group characteristics for interaction. Multi-agent RPAs involve more complex interactions, with social simulacra capturing interpersonal dynamics within small, predefined groups, and synthetic societies modeling large-scale collective decision-making and societal structures.

\subsection{Evaluation of RPAs}
Existing surveys on the evaluation of RPAs~\cite{gao2024large, chen2024from, tseng-etal-2024-two, chen2024oscars, mou2024individual} provide a unified classification of RPA evaluation metrics from the perspective of evaluation approaches. However, they lack a comprehensive and consistent taxonomy for versatile evaluation metrics, leading to arbitrary metrics selection in practices.

Prior works~\cite{gao2024large,mou2024individual} categorize RPA evaluations into three types: automatic evaluations, human-based evaluations, and LLM-based assessments. Automatic evaluations are efficient and objective, but lack context sensitivity, failing to capture nuances like persona consistency. Human-based evaluations provide deep insight into character alignment and engagement, but they are costly, less scalable, and prone to subjectivity. LLM-based evaluations are automatic and offer scalability and speed, but may not always align with human judgments.

The classification of evaluation metrics in prior works varies significantly, leading to inconsistency and ambiguity. For instance, \citet{gao2024large} focuses on realness validation and ethics evaluation, whereas \citet{chen2024from} differentiates between character persona and individualized persona. Furthermore, \citet{chen2024oscars} classifies evaluation into conversation ability, role-persona consistency, role-behavior consistency, and role-playing attractiveness, which partially overlap with \citet{mou2024individual}'s individual simulation and scenario evaluation. These discrepancies indicate a lack of standardized taxonomy, making it difficult to compare results across studies and select appropriate evaluation metrics for specific applications.

While existing surveys offer different taxonomies of RPA evaluation, they do not provide concrete evaluation design guidelines. Our work addresses this gap by proposing a structured framework that systematically links evaluation metrics to RPA attributes and real-world applications.

\section{Method}

We conduct a systematic literature review to address our research question. Following prior method~\cite{nightingale2009guide}, we aim to identify relevant research papers on RPAs and provide a comprehensive summary of the literature. We selected four widely used academic databases: Google Scholar, ACM Digital Library, IEEE Xplore, and ACL Anthology. These databases encompass a broad spectrum of research across AI, human-computer interaction, and computational linguistics. Given the rapid advancements in LLM research, we included both peer-reviewed and preprint studies (e.g., from arXiv) to capture the latest developments. Below, we detail our paper selection and annotation process.

\begin{table*}[t]
\small

\caption{Definition and examples of six agent attributes.}
\resizebox{\textwidth}{!}{%
\begin{tabular}{@{}p{0.21\textwidth}p{0.45\textwidth}p{0.33\textwidth}@{}}
\toprule
\textbf{Agent attributes}     & \textbf{Definition}     & \textbf{Examples} \\ 
\midrule
Activity History        & A record of past actions, behaviors, and engagements, including schedules, browsing history, and lifestyle choices. & Backstory, plot, weekly schedule, browsing history, social media posts, lifestyle       \\ 
Belief and Value        & The principles, attitudes, and ideological stances that shape an individual's perspectives and decisions.           & Stances, beliefs, attitudes, values, political leaning, religion                            \\ 
Demographic Information & Personal identifying details such as name, age, education, career, and location.                                    & Name, appearance, gender, age, date of birth, education, location, career, household income \\ 
Psychological Traits    & Characteristics related to personality, emotions, interests, and cognitive tendencies.                              & Personality, hobby and interest, emotional                                                  \\ 
Skill and Expertise     & The knowledge level, proficiency, and capability in specific domains or technologies.                             & Knowledge level, technology proficiency, skills                                            \\ 
Social Relationships & The nature and dynamics of interactions with others, including roles, connections, and communication styles.        & Parenting styles, interactions with players                                                \\ 
\bottomrule
\end{tabular}
}
\vspace{-1em}
\label{attr_def}
\end{table*}

\subsection{Literature Search and Screening Method}

\begin{figure}
    \includegraphics[width=\linewidth]{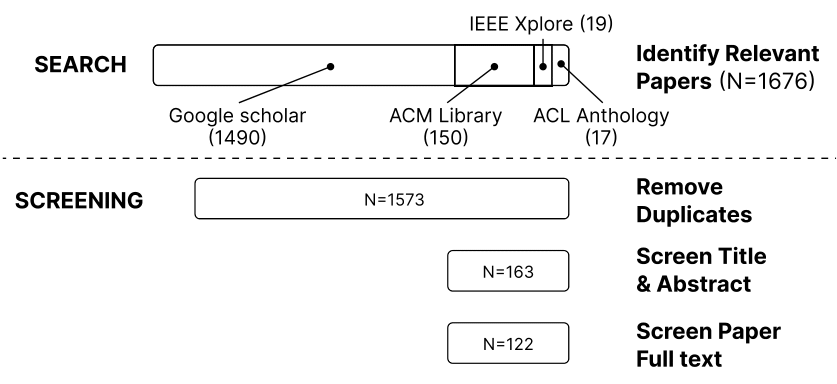}
    \caption{Screening process of literature review. We initially retrieved $1,676$ papers published between 2021 and 2024, and narrowed down to $122$ final selections.}
    \vspace{-1em}
    \label{fig:prisma}
\end{figure}

Our literature review focuses on LLM agents that role-play human behaviors, such as decision-making, reasoning, and deliberate actions. We specifically focus on studies where LLM agents demonstrate the ability to simulate human-like cognitive processes in their objectives, methodologies, or evaluation techniques. To ensure methodological rigor, we define explicit inclusion and exclusion criteria (Tab.~\ref{tab:criteria} in Appendix~\ref{tab: inclusion and exclusion criteria}). 

The inclusion criteria require that an LLM agent in the study exhibits human-like behavior, engages in cognitive activities such as decision-making or reasoning, and operates in an open-ended task environment. We excluded studies where LLM agents primarily serve as chatbots, task-specific assistants, evaluators, or agents operating within predefined and finite action spaces. Additionally, studies focusing solely on perception-based tasks (e.g., computer vision or sensor-based autonomous driving) without cognitive simulation were also excluded.

Using this scope, we searched four databases using the query string provided in Appendix~\ref{query string}, retrieving $1,676$ papers published between January 2021 to December 2024. After removing duplicates, $1,573$ unique papers remained. Two authors independently screened the paper titles and abstracts based on the inclusion criteria. If at least one author deemed a paper relevant, it proceeded to full-text screening, where two authors reviewed the paper in detail and resolved any disagreements through discussion (Fig.~\ref{fig:prisma}). The final set of selected studies comprised $122$ publications.

\subsection{Paper Annotation Method}
Our team followed established open coding procedures \cite{brod2009qualitative} to conduct an inductive coding process to identify key themes. Three co-authors with extensive experience in LLM agents (``annotators,'' hereinafter) collaboratively annotated the papers on three dimensions: \textbf{agent attributes}, \textbf{task attributes}, and \textbf{evaluation metrics}. 

To ensure consistency, two annotators independently annotated the same 20\% of articles and then held a meeting to discuss and refine an initial set of categories for the three dimensions. After reaching a consensus, each annotator annotated half of the remaining papers and cross-validated the other half annotated by the other annotator. Once the annotations were completed, a third annotator reviewed the coded data and identified potential discrepancies. 
Any discrepancies were discussed among the annotators to ensure consistency until disagreements were resolved, ensuring reliability and validity through an iterative refinement process.
\section{Survey Findings}

Building on the annotated data, we systematically categorized agent attributes, task attributes, and evaluation metrics. 
We then present a structured RPA evaluation design guideline, outlining how to select appropriate evaluation metrics based on agent and task attributes.

\begin{table*}[ht]
\small
\caption{Definition of seven task attributes.}
\begin{tabular}{@{}ll@{}}
\toprule
\textbf{Task attributes}         & \textbf{Definition}                                                                                  \\ \midrule
Simulated Individuals    & Simulating specific individuals or groups, such as users and participants.                                                                                                  \\ 
Simulated Society        & Simulating social interactions, such as cooperation, competition, and communication. \\ 
Opinion Dynamics         & Simulating political views, legal perspectives, and social media content.                                                                                                                  \\ 
Decision Making          & Simulating decision-making of stakeholders in investment, public policies, or games.                                                                                            \\ 
Psychological Experiments & Simulating human traits, including personality, ethics, emotions, and mental health.                                                                                                                             \\ 
Educational Training     & Simulating teachers and learners to enable personalized teaching and accommodate learner needs.                      \\ 
Writing                  & Simulating readers or characters to support character development and audience understanding.                             \\  \bottomrule
\end{tabular}
\label{task_def}
\end{table*}


\begin{table*}[]
\small
\caption{Definitions and examples of seven evaluation metric categories.}
\begin{tabular}{@{}p{0.22\textwidth}p{0.56\textwidth}p{0.17\textwidth}@{}}
\toprule
\textbf{Evaluation Metrics}                   & \textbf{Definitions}          & \textbf{Examples}                                                                                                                    \\ \midrule
Performance                 & Assess RPAs' effectiveness in task execution and outcomes.  & Prediction accuracy                                                                                     \\ 
Psychological               & Measure human psychological responses to RPAs and the agents' self-awareness and emotional state.      & Big Five Invertory                                          \\ 
External Alignment          & Evaluate how closely RPAs align with external ground truth or human behavior and judgments.  & Alignment between model and human                                                  \\ 
Internal Consistency        & Assess coherence between an RPA’s predefined traits (e.g., personality), contextual expectations, and behavior. & Personality-behavior alignment                                  \\ 
Social and Decision-Making  & Analyze RPAs' social interactions and decision-making, including their effects on negotiation, societal welfare, markets, and social dynamics. & Social Conflict Count  \\ 
Content and Textual        & Evaluate the quality, coherence, and diversity of RPAs' text, including semantic understanding, linguistic style, and engagement.  & Content similarity               \\ 
Bias, Fairness, and Ethics  & Assess biases, extreme or unbalanced content, or stereotyping behavior.   & Factual error rate                                                                        \\ \bottomrule
\end{tabular}
\label{metrics_def}
\end{table*}

\subsection{Agent Attributes}
We identified six categories of agent attributes, as shown in Tab.~\ref{attr_def}. 
\textit{Activity history} refers to an agent's longitudinal behaviors, such as browsing history~\cite{10.1145/3613904.3642363} or social media activity~\cite{Navarro2024DesigningRE}. 
\textit{Belief and value} encompass the principles, attitudes, and ideological stances that shape an agent's perspectives, including political leanings~\cite{Mou2024UnveilingTT} or religious affiliations~\cite{lv2024coggpt}. 
\textit{Demographic information} includes personal details such as name, age, education, location, career status, and household income. 
\textit{Psychological traits} include an agent's personality~\cite{NEURIPS2023_21f7b745}, emotions, and cognitive tendencies~\cite{castricato2024personareproducibletestbedpluralistic}. 
\textit{Skill and expertise} describe an agent's knowledge and proficiency in specific domains, such as technology proficiency or specialized professional skills. 
Lastly, \textit{social relationships} define the social interactions, roles, and communication styles between agents, including aspects like parenting styles~\cite{ye2024simulating} or relationships between players~\cite{ge2024scaling}.

\subsection{Task Attributes}
We identified seven key types of RPA downstream task attributes (Tab.~\ref{task_def}). These tasks fall into two broad categories: those that use simulation as a research goal and those that use simulation as a tool to support specific research domains.

Among them, simulated individuals and simulated society primarily use simulation as the research goal. 
\textit{Simulated individuals} involve modeling specific individuals or groups, such as end-users~\cite{chen2024seeing}, to study their behaviors and interactions in a controlled setting. 
\textit{Simulated Society} focuses on social interactions, including cooperation~\cite{bouzekri2024chatgpt}, competition~\cite{wu2024shall}, and communication~\cite{mishra-etal-2023-e}, aiming to explore emergent social dynamics.\looseness=-1

In contrast, the other task attributes employ simulation as a means to serve specific research domains. 
\textit{Opinion dynamics} entails simulating political views~\cite{neuberger2024sauce}, legal perspectives~\cite{chen2024agentcourt}, and social media discourse~\cite{liu2024tiny} to analyze the formation and evolution of opinions. \textit{Decision making} addresses the decision-making processes of stakeholders in investment~\cite{sreedhar2024simulating} and public policy~\cite{ji2024srap}, providing insights into strategic behaviors. \textit{Psychological experiments} explore human traits such as personality~\cite{bose2024assessing}, ethics~\cite{lei2024fairmindsim}, emotions~\cite{zhao2024esc}, and mental health~\cite{de2025introducing}, using simulated scenarios to study cognitive and behavioral responses. 
\textit{Educational training} supports personalized learning by simulating teachers and learners, enhancing pedagogical approaches and adaptive education systems~\cite{Liu2024PersonalityawareSS}. Finally, \textit{writing} involves modeling readers or characters to facilitate character development~\cite{10.1145/3613904.3642406} and audience engagement~\cite{choi2024proxona}, contributing to storytelling and content generation research.

\begin{figure*}[t]
    \centering
    \includegraphics[width=0.9\linewidth]{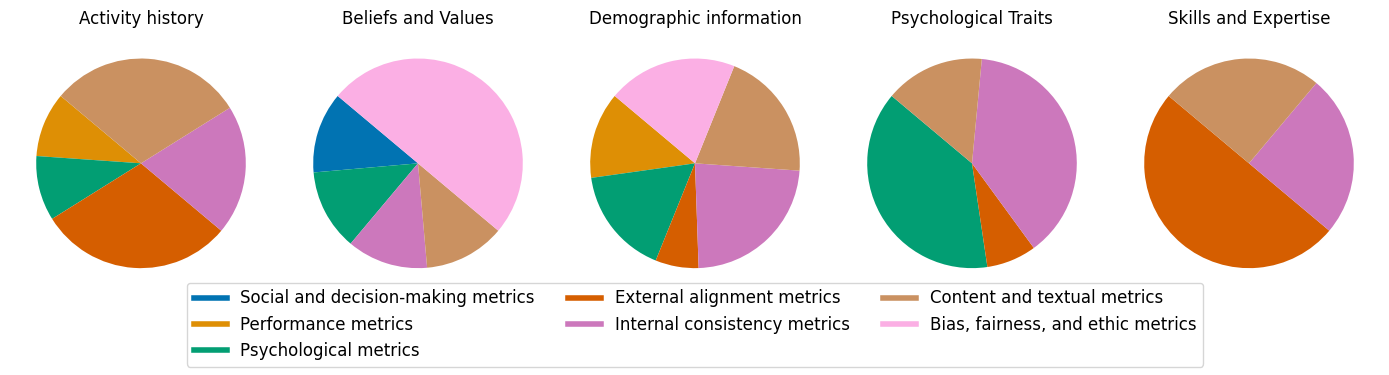}
    \caption{Proportional distribution of agent-oriented metrics across different agent attributes.}
    \label{fig:pie-chart-agent-oriented}
\end{figure*}

\begin{figure*}
    \centering
    \includegraphics[width=\linewidth]{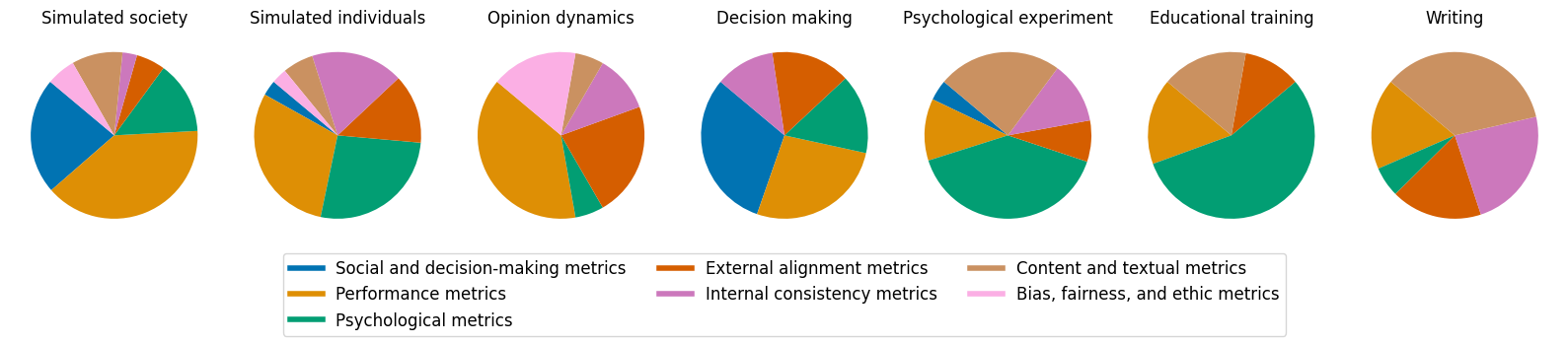}
    \caption{Proportional distribution of task-oriented metrics across different task attributes.}
    \label{fig:pie-chart-task-oriented}
\end{figure*}

\subsection{Agent- and Task-Oriented Metrics}
We derived seven categories of evaluation metrics (Tab.~\ref{metrics_def}) that are shared by agent- and task-oriented metrics despite differences in the specific metrics. 

\begin{table}[]
\small
\resizebox{\linewidth}{!}{%
\begin{tabular}{@{}p{0.32\linewidth}p{0.7\linewidth}@{}}
\toprule
\textbf{Agent Attributes}        & \textbf{Top 3 Agent-Oriented Metrics}                                                           \\ \midrule
Activity History        & External alignment metrics, internal consistency metrics, content and textual metrics  \\
Belief and Value        & Psychological metrics, bias, fairness, and ethics metrics                              \\
Demographic Info. & Psychological metrics, internal consistency metrics, external alignment metrics        \\
Psychological Traits    & Psychological metrics, internal consistency metrics, content and textual metrics       \\
Skill and Expertise     & External alignment metrics, internal consistency metrics,  content and textual metrics \\
Social Relationship     & Psychological metrics, external alignment metrics, social and decision-making metrics  \\ \bottomrule
\end{tabular}
}
\caption{Top 3 frequently used agent-oriented metrics for each agent attribute}
\label{tab: top3-agent}
\vspace{-1em}
\end{table}
\begin{table}[t]
\small
\resizebox{\linewidth}{!}{%
\begin{tabular}{@{}p{0.37\linewidth}p{0.63\linewidth}@{}}
\toprule
\textbf{Task Attributes}          & \textbf{Top 3 Task-Oriented Metrics}                                                             \\ \midrule
Simulated Individuals    & Psychological, performance, and internal consistency metrics                            \\
Simulated Society        & Social and decision-making metrics, performance metrics, and psychological metrics      \\
Opinion Dynamics         & Performance metrics, external alignment metrics, and bias, fairness, and ethics metrics \\
Decision Making          & Social and decision-making, performance, and psychological metrics                      \\
Psychological Experiment & Psychological, content and textual, and performance metrics                             \\
Educational Training     & Psychological, performance, and content and textual metrics                             \\
Writing                  & Content and textual, psychological, and performance metrics                             \\ \bottomrule
\end{tabular}
}
\caption{Top 3 frequently used task-oriented metrics for each task attribute}
\label{tab: top3-task}
\vspace{-1em}
\end{table}

\textbf{Agent-oriented metrics} focus on intrinsic, task-agnostic properties that define an RPA's essential ability, such as underlying reasoning, consistency, and adaptability. 
These include \textit{performance} metrics like memorization, \textit{psychological} metrics such as emotional responses measured via entropy of valence and arousal, and \textit{social and decision-making} metrics like social value orientation. 
Additionally, agent-oriented evaluations emphasize \textit{internal consistency} metrics (e.g., consistency of information across interactions), \textit{external alignment} metrics (e.g., hallucination detection), and \textit{content and textual} metrics such as clarity. 
These evaluations ensure logical coherence, factual accuracy, and alignment with expected behavioral and cognitive frameworks, independent of any specific task. 

\textbf{Task-oriented metrics} evaluate an RPA's effectiveness in performing specific downstream tasks, focusing on task-related aspects such as accuracy, consistency, social impact, and ethical considerations. 
\textit{Performance} measures how well RPAs execute designated tasks, such as prediction accuracy. 
\textit{Psychological} metrics assess human psychological responses to RPAs, including self-awareness and emotional states; for example, the Big Five Inventory. 
\textit{External alignment} evaluates how closely RPAs align with external ground truth or human behavior; for instance, alignment between model and human. 
\textit{Internal consistency} ensures coherence between an RPA’s predefined traits, contextual expectations, and behavior; for example, personality-behavior alignment. 
\textit{Social and decision-making} metrics analyze RPAs’ influence on negotiation, societal welfare, and social dynamics; for instance, the social conflict count. 
\textit{Content and textual quality} focuses on the coherence, linguistic style, and engagement of RPAs’ generated text, such as content similarity. 
Lastly, \textit{bias, fairness, and ethics} metrics examine biases, extreme content, or stereotypes; for instance, the factual error rate.
Together, these seven metrics provide a comprehensive framework for evaluating RPAs' task performance and broader impact.

\subsection{RPA Evaluation Design Guideline}
Building on our previous classification of agent attributes, task attributes, and evaluation metrics, we observed that both agent design and evaluation can be broadly divided into two categories: \textbf{agent-oriented} and \textbf{task-oriented}. This distinction led us to investigate patterns between agent design and evaluation, aiming to develop systematic guidelines for selecting evaluation metrics in future research.

\paragraph{Step 1. Selecting Agent-Oriented Metrics Based on Agent Attributes}
We analyzed the distribution of agent attributes and agent-oriented metrics, as illustrated in Fig.~\ref{fig:pie-chart-agent-oriented}. Our analysis reveals that, for each agent attribute, the top three categories of agent-oriented metrics account for the majority of all metric types. Based on this observation, our first guideline recommends selecting agent-oriented metrics according to agent attributes. Specifically, we suggest referring to Tab.~\ref{tab: top3-agent} to identify the top three corresponding metrics. For instance, for Activity History, the recommended metrics are external alignment, internal consistency, and content and textual metrics. Likewise, for Beliefs and Values, the most relevant choices are psychological metrics and bias, fairness, and ethics metrics.
In particular, there are no established agent-oriented evaluation metrics for social relationships. Based on Social Exchange Theory~\cite{cropanzano2005social}, which explains relationship formation through reciprocal interactions and resource exchanges, we propose assessing social relationships with psychological metrics, external alignment metrics, and social and decision-making metrics.

\paragraph{Step 2: Selecting Task-Oriented Metrics Based on Task Attributes}
Additionally, we analyzed the distribution of task attributes and task-oriented metrics, as shown in Fig.~\ref{fig:pie-chart-task-oriented}. Consistent with our previous findings, we observed that for each category of task attributes, the top three task-oriented metrics account for the vast majority of all metrics. Based on this, our second guideline recommends selecting task-oriented metrics according to task attributes. Specifically, we suggest referring to Tab.~\ref{tab: top3-task}  to identify the top three corresponding metrics. For instance, for the \textit{Simulated Society} task, the recommended metrics are social and decision-making, performance, and psychological metrics. Similarly, for the \textit{Opinion Dynamics} task, the most relevant choices are performance, external alignment, bias, fairness, and ethics metrics.

However, these two steps should not be treated as one-time decisions. As the agent design process evolves, evaluation results may prompt adjustments to the attributes of the agent and the task, thereby influencing the selection of evaluation metrics. Therefore, this two-step evaluation guideline should be used iteratively to ensure that the evaluation remains adaptive to changing agent capabilities and task requirements. This iterative approach enhances the reliability, relevance, and robustness of RPA evaluation experiments.

\section{Case Study: How to Use RPA Design Guideline to Select Evaluation Metrics}

We present \textbf{two case studies} to illustrate how following our evaluation guidelines leads to the selection of a comprehensive set of evaluation metrics, while significant deviations may result in incomplete evaluation. By adopting the perspective of the original authors, we compare the evaluation outcomes resulting from adhering to or deviating from the RPA evaluation guidelines.

\subsection{A Good Example: \textit{Generative Agents: Interactive Simulacra of Human Behavior}}
\label{sec_good_example}

As shown in Fig.~\ref{fig:teaser}, \citet{park2023generative} designed agents with demographic information, action history, and social relationships to create an interactive artificial society. 
Their evaluation methods are in line with the structured selection process proposed in our survey. Since no established agent-oriented evaluation metrics exist for social relationships, they focused on demographic information and action history. Referring to Fig.~\ref{fig:pie-chart-agent-oriented}, they identified four relevant metric categories: Content and textual metrics, Internal consistency metrics, External alignment metrics, and Psychological metrics. Based on Tab.~\ref{tab: long_table_agent_metrics_src} in Appendix~\ref{sec: metrics glossary}, they selected five specific evaluation metrics: Self-knowledge (Content and textual, Internal consistency), Memory and Plans (Internal consistency), Reactions (External alignment), and Reflections (Psychological).

For task-oriented metrics, they determined that the agents’ downstream tasks aligned with \textit{simulated society} and designed the evaluation metrics that are aligned with the top three most relevant metric types reported in Fig.~\ref{fig:pie-chart-task-oriented}.
As shown in Tab.~\ref{tab: long_table_task_metrics_src} in Appendix~\ref{sec: metrics glossary}, they selected four evaluation metrics: Response accuracy (Performance), Relationship formation (Psychological), Information diffusion and Coordination (Social and decision-making). By systematically aligning evaluation metrics with agent attributes and task objectives, this approach ensured a comprehensive and meaningful assessment.

\subsection{A Flawed Example: \textit{A Generative Social World for Embodied AI}}
\label{sec_bad_example}

A flawed example is presented in Appendix~\ref{appendix:bad_example} Fig.~\ref{fig:bad example}, which is an ICLR submission, and the reviews are publicly available on OpenReview.
The authors
developed agents with demographic attributes, action history, psychological traits, and social relations for route planning and election campaigns. However, their evaluation deviated significantly from our RPA evaluation design guidelines.

Despite designing agents with clear attributes, they did not include any agent-oriented evaluation metrics. For task-oriented metrics, they identified tasks related to Opinion Dynamics and Decision-Making, which should have been evaluated using five key categories: Performance metrics, Psychological metrics, External alignment metrics, Social and decision-making metrics, and Bias, fairness, and ethics metrics. Instead, their evaluation relied solely on Arrival rate, Time, and Alignment between campaign strategies, leading to an incomplete assessment. This omission resulted in criticism from reviewers, as one noted: \textit{``The paper performs almost no quantitative experiments... This actually shows that the benchmark cannot cover too many current research methods, which is the biggest weakness of the paper.''}

\begin{figure}[t]
    \centering
    \includegraphics[width=\linewidth]{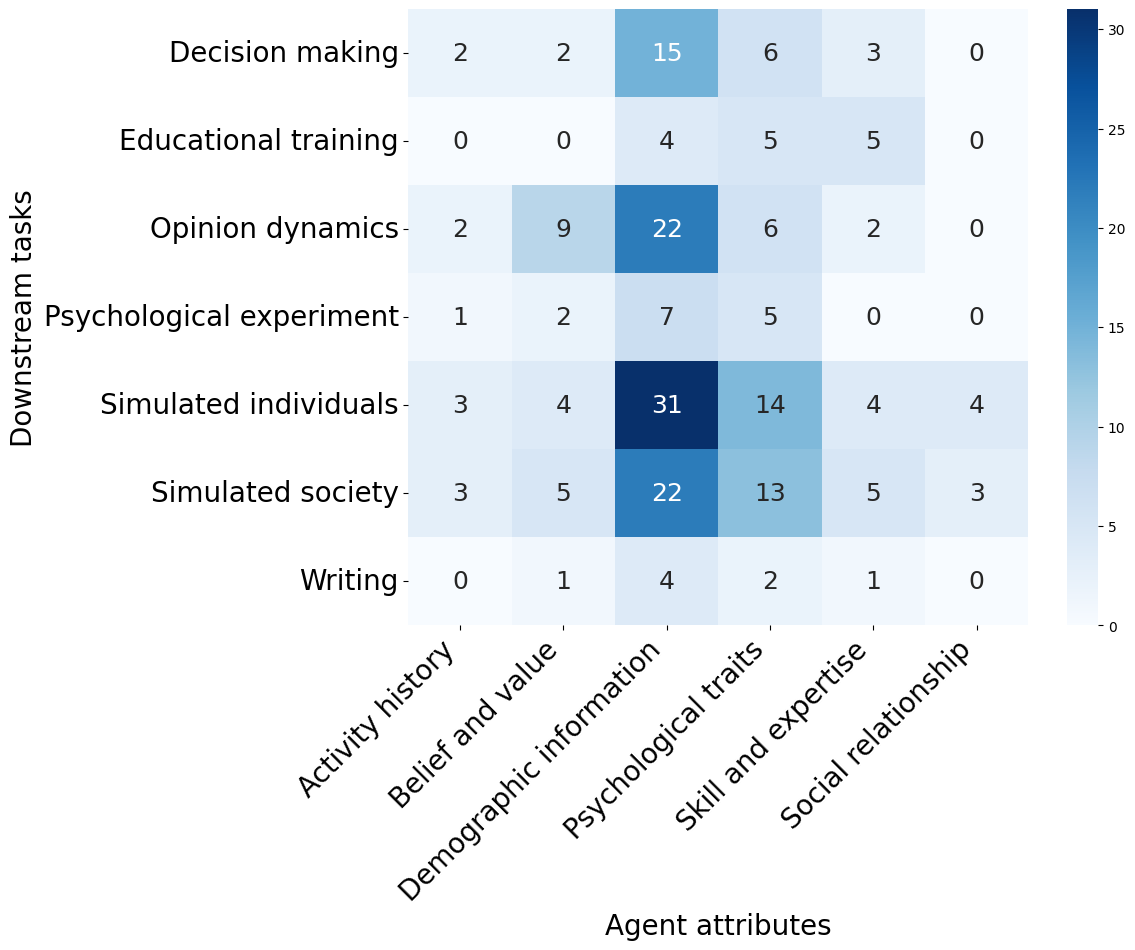}
    \caption{Relationships between agent attributes and downstream tasks. The numbers in the heatmap represent the paper counts.}
    \label{fig:heatmap}
\end{figure}

\section{Relationships Between Agent Attributes and Downstream Tasks}

Both agent attributes and downstream task attributes play a crucial role in selecting appropriate RPA evaluation metrics. Researchers predefine these factors when designing and evaluating RPAs, yet their interrelation remains an open question.  In this section, we analyze how agent attributes correspond to different downstream tasks, uncovering several recurring patterns (Fig.~\ref{fig:heatmap}).

Demographic information and psychological traits are fundamental across all downstream tasks. Whether in decision-making, opinion dynamics, or simulated environments, these attributes consistently shape RPA design. As shown in Fig.~\ref{fig:heatmap}, they are the most frequently incorporated factors, underscoring their central role in modeling agent behavior across diverse applications.

For tasks where simulation itself is the primary objective, such as Simulated Individuals and Simulated Society, the selection of agent attributes becomes broader. In addition to demographic and psychological factors, these tasks frequently incorporate skills, expertise, and social relationships, reflecting the need for richer agent representations to capture complex social and individual interactions. By contrast, tasks that use simulation as a means to study specific research fields tend to prioritize certain agent attributes. For instance, in Opinion Dynamics, beliefs and values play a distinctive role, as they directly influence how agents interact and form opinions. Similarly, tasks related to Educational Training and Writing exhibit a different pattern, emphasizing skills and expertise over broad demographic or psychological considerations.

In contrast, attributes such as activity history and social relationships receive significantly less emphasis across tasks. This raises a question: is their impact inherently limited, or are they simply underexplored in current RPA applications?

Overall, these findings highlight the nuanced interplay between agent attributes and downstream tasks. While demographic information and psychological traits are universally relevant, attributes like beliefs and values gain importance in specific contexts. At the same time, the relative absence of activity history and social relationships in current evaluations presents an open research question, particularly in scenarios requiring long-term modeling and complex social interactions.


\section{Discussion}
\label{sec:discussion}

\subsection{RPA: an Algorithm v.s. a System}

Unlike traditional algorithmic innovations in NLP, the design of RPAs can not only support technical innovations to improve LLMs' humanoid capabilities but also enable RPA-based simulation systems for practical benefits.
For instance, from the perspective of psychology, RPAs support the exploration of human cognitive and behavioral activities in controlled yet highly scalable experiments, even in hypothetical scenarios.
In social science, RPAs can deployed as proxies or pilot experiments to analyze and audit social systems, power dynamics, and human societal behaviors at scale.
For the machine learning community, RPAs shed light on dynamic and human-centered model evaluations that are aligned with real-world scenarios by incorporating human and societal factors into consideration.
Last but not least, HCI researchers are particularly intrigued by the implications of RPA systems that can provide personalized assistance with human-centered applications in various sectors, such as medicine, healthcare, and education. 

Nevertheless, RPAs' capability and flexibility are a double-edged sword; they not only have the potential to bring benefits to stakeholders but also expose potential risks and even harm if not responsibly designed.
To what extent do RPAs' responses align with genuine human cognitive activities, whether the cultural, linguistic, and contextual biases learned from the training data of LLMs impact predicted behaviors, and how to ensure RPAs' robustness and consistency under different scenarios, are critical but under-explored challenges for both technical developers and system designers.

As a result, the design of RPAs should incorporate system design considerations while advancing technical explorations.
For instance, RPA design should focus on target users from the very beginning of system design, emphasize the diversity of user backgrounds and perspectives, and iteratively refine the system, as suggested by \citet{gould1985designing} and \citet{shneiderman2010designing} in established design guidelines for system usability.
Nevertheless, differences in cultural norms, linguistic subtleties, and domain-specific knowledge can introduce variability in how RPAs are designed and perceived.
Designers and developers must focus on a balance between generalization and specificity to ensure RPAs are both adaptable and effective across a wide range of scenarios. 

\subsection{The Design of RPA Persona}

One of RPAs' key strengths is their ability to adapt to diverse personas, tasks, and environments. But how can RPA personas be designed to ensure that LLMs faithfully and believably reflect the agents' cognitive behaviors within a given task?
Persona descriptions must strike a careful balance between intrinsic agent characteristics and contextual factors, ensuring thoughtful consideration of both the agents' intrinsic characteristics and the contextual information of the specific environments for which the agents are designed.



The \textit{intrinsic characteristics} of RPAs, such as their personal characteristics, education experience, domain expertise, emotional expressiveness, and decision-making processes, must be \textit{aligned with the purpose} of the applications of RPAs.
For example, an RPA designed for psychological experiments should prioritize cognitive characteristics like personality and empathy ability, whereas an RPA developed for economic simulations might emphasize negotiation tactics, competitive reasoning, and adaptability to changing conditions.

On the other hand, \textit{contextual information}, such as task- and scenario-specific details, factors, and specifications, is equally critical in shaping the behaviors of RPAs. 
In healthcare applications, for instance, RPAs may simulate caregivers' emotional responses to patients' changing health status but still operate under clinical protocols, such as the ICU visitor rules.
The granularity and fidelity of contextual information heavily influence the believability and effectiveness of the agents' behaviors.


\subsection{The Challenges of RPA Evaluation}

The versatility of RPAs, which allows them to function in diverse roles and contexts, makes it infeasible to have a ``one-solution-fits-all'' evaluation metric for systematically evaluate RPAs both within and across tasks and user scenarios.
One major difficulty lies in designing and determining task-oriented and agent-oriented evaluation metrics. 
Despite our work recommending an RPA evaluation design guideline based on a comprehensive review of the literature, existing evaluation metrics may not be sufficient to measure the performance of RPAs for different domain-specific applications.


The diversity of user scenarios further exacerbates the evaluation challenge. 
Different tasks may prioritize different aspects of RPAs, making it difficult to develop a one-size-fits-all evaluation framework.
For instance, RPAs designed for psychological research focus on believable emotional responses, whereas RPAs for policymaking simulations underscore robustness to policy changes.

Moreover, cross-task evaluations pose significant challenges due to inconsistencies in how metrics are designed and applied across studies. 
The lack of standardized evaluation criteria complicates systematic benchmarking in RPA development and impedes interdisciplinary collaboration.

Addressing these challenges will require the development of systematic, multi-faceted evaluation frameworks that can accommodate the diverse applications and capabilities of RPAs while providing consistency and comparability across studies.

\section{Conclusion}
RPA evaluation lacks consistency due to varying tasks, domains, and agent attributes. Our systematic review of $1,676$ papers reveals that task-specific requirements shape agent attributes, while both task characteristics and agent design influence evaluation metrics. By identifying these interdependencies, we propose guidelines to enhance RPA assessment reliability, contributing to a more structured and systematic evaluation framework.
\section*{Limitations}

RPAs are rapidly evolving and have widespread applications across various domains. While we aim to comprehensively review existing literature, we acknowledge certain limitations in our scope.
First, our review may not encompass all variations of RPA evaluation approaches across different application domains. Second, new research published after December 2024 is not included in our analysis. As a result, our work does not claim to exhaustively cover all potential evaluation metrics. Instead, our goal is to provide a structured framework and actionable guidelines to help future researchers design more systematic and consistent RPA evaluations, even as the field continues to evolve.

\section*{Ethics Statement}

Our work focuses on summarizing and analyzing the evaluation of RPAs, which we believe will be valuable to researchers in AI, HCI, and related fields such as psychological simulation, educational simulation, and economic simulation. We have taken care to ensure that this survey remains objective and balanced, neither overestimating nor underestimating trends. We do not anticipate any ethical concerns that arise from the research presented in this paper.

\bibliography{bibliography}
\clearpage
\appendix

\section{Inclusion and Exclusion Criteria}
\label{tab: inclusion and exclusion criteria}

We summarize the inclusion and exclusion criteria in Table~\ref{tab:criteria}. Briefly, the \textbf{Inclusion Criteria (IC)} ensure that the reviewed studies focus on LLM agents exhibiting human-like behavior—either implicitly (e.g., preference or behavioral patterns) or explicitly (e.g., emotions or personality)—along with key cognitive processes such as reasoning and decision-making. Moreover, an open-ended action space and the capacity to tackle multifaceted tasks are essential attributes for inclusion.

By contrast, the \textbf{Exclusion Criteria (EC)} eliminate studies employing LLMs purely as chatbots, single-purpose systems, or evaluation tools, rather than as agents mimicking human cognition. Likewise, if the LLM agents are restricted to fixed, close-ended tasks or limited to algorithmic optimization without simulating cognitive processes, they fall outside the scope of this work.
\begin{table}[t]
\small
\caption{Inclusion and exclusion criteria.}
\label{tab:criteria}
\begin{tabularx}{\columnwidth}{lX}
    \toprule
    \multicolumn{2}{l}{\textbf{Inclusion Criteria (IC)}} \\
    \midrule
        IC-1 & The LLM agents in the paper simulate humanoid behavior with implicit personality (e.g., preference and behavior pattern) or explicit personality (e.g., emotion or characteristics).  \\
        IC-2 & The LLM agents in the paper have cognitive activities such as decision-making, reasoning, and planning. \\
        IC-3 & The LLM agents in the paper are capable of completing complicated and general tasks. \\
        IC-4 & The LLM agents' action set in the paper is neither predefined nor finite. \\
    \midrule
    \multicolumn{2}{l}{\textbf{Exclusion Criteria (EC)}} \\
    \midrule
        EC-1 & The study does not employ LLM agents for simulation purposes but rather uses them as chatbots, task-specific agents, or evaluators. \\
        EC-2 & The paper's research objectives, methodologies, and evaluations are not focused on simulating human-like behavior with LLM agents, but rather on optimizing LLM algorithms. \\
        EC-3 & The study primarily investigates the perception or action capabilities of LLM agents without simulating the cognitive process. \\
        EC-4 & The LLM agents are restricted to handling specific, close-ended tasks. \\
        EC-5 & The LLM agents' actions are either predefined or limited. \\
    \bottomrule
\end{tabularx}
\end{table}

\section{Query String}
\label{query string}

We employed the following query to guide our literature retrieval process:

\begin{quote}
\texttt{(“large language model” OR LLM) AND (agent OR persona OR "human digital twin" OR simulacra) AND (simulat* OR generat* OR eval*) AND “human behavior” AND cognit*}
\end{quote}

This query was designed to capture a broad spectrum of studies on large language models that simulate or replicate human-like behavior. It combines keywords related to LLM agents (\emph{LLM}, \emph{persona}, \emph{simulacra}), their capabilities (\emph{simulat*}, \emph{generat*}, \emph{eval*}), and the focus on cognitively grounded human behavior (\emph{cognit*}). This ensures that the resulting literature is relevant to our exploration of how LLM-based systems can mimic or exhibit human-like cognition and behavior patterns.

\section{Evaluation Approach Usage for Agent- and Task-Oriented Metrics}
\begin{figure}[t]
    \centering
    \includegraphics[width=\linewidth]{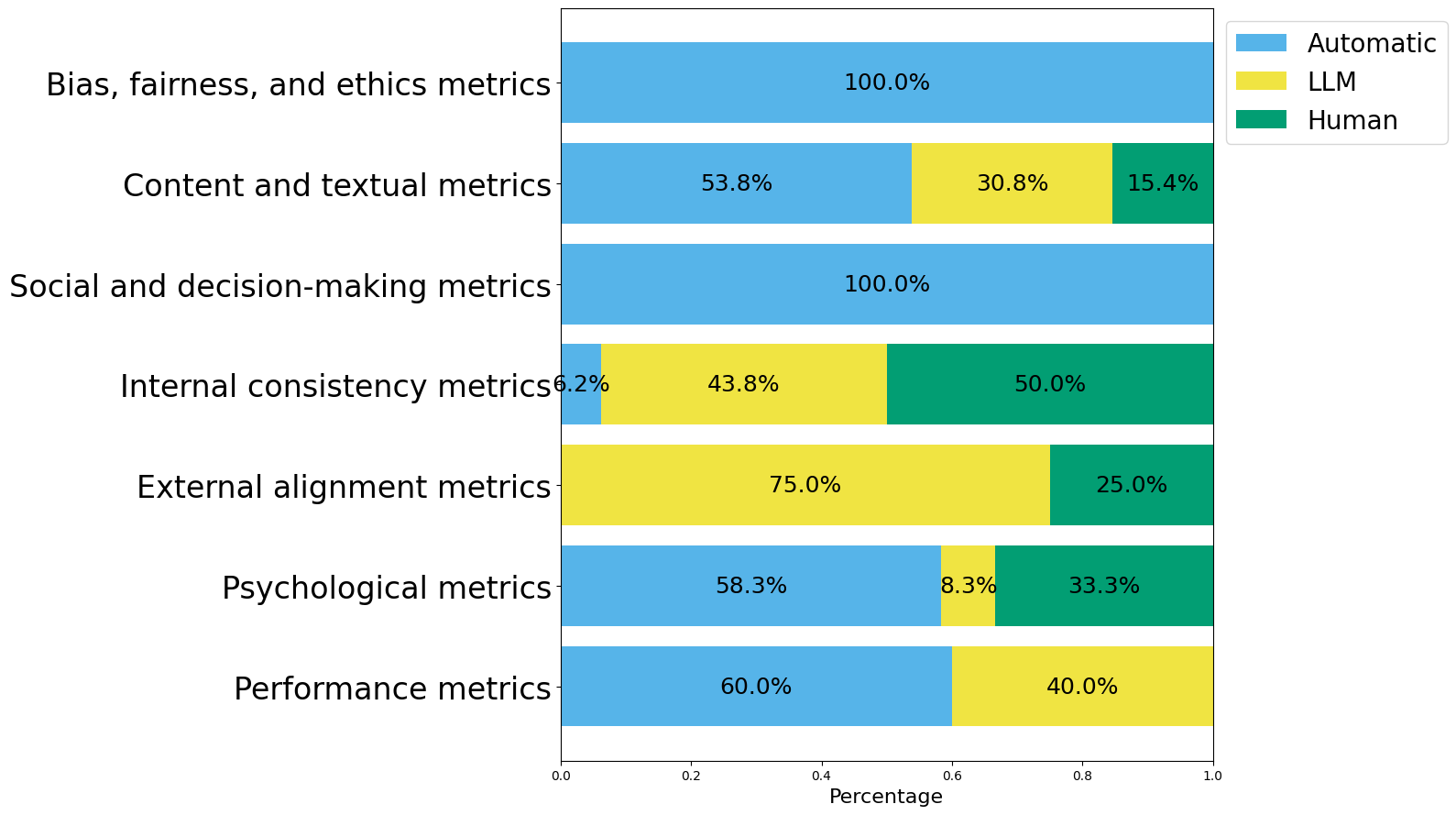}
    \caption{Usage ratio of evaluation approaches for each category of agent-oriented metrics.}
    \label{fig:stacked-bar-agent-oriented}
\end{figure}

\begin{figure}[t]
    \centering
    \includegraphics[width=\linewidth]{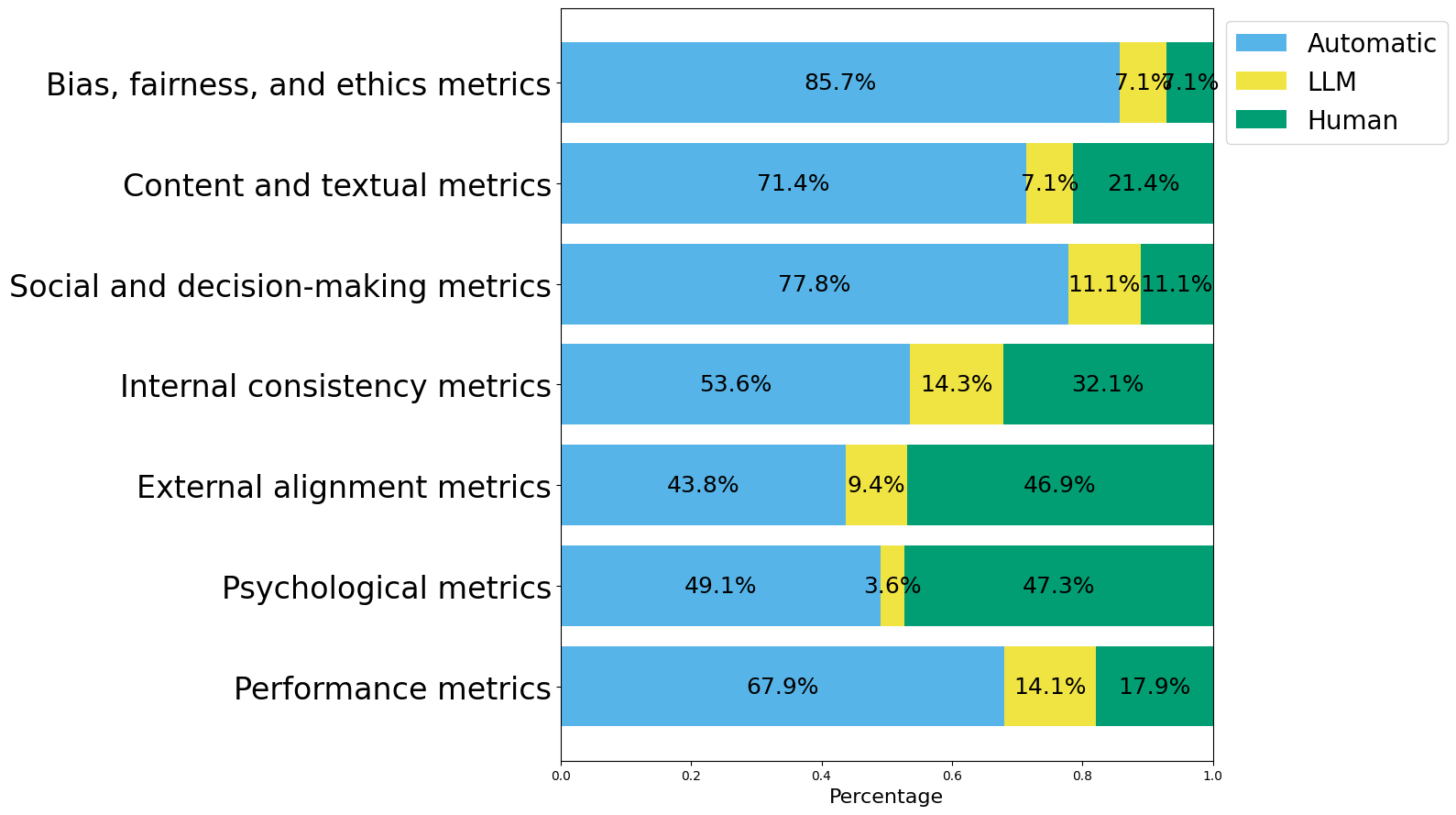}
    \caption{Usage ratio of evaluation approaches for each category of task-oriented metrics.}
    \label{fig:stacked-bar-task-oriented}
\end{figure}

We present a breakdown of evaluation approach usage by agent-oriented metrics (Fig.~\ref{fig:stacked-bar-agent-oriented}) and task-oriented metrics (Fig.~\ref{fig:stacked-bar-task-oriented}).

\section{Case Study: Flawed Example}
\label{appendix:bad_example}

\begin{figure}[t]
    \centering
    \includegraphics[width=\linewidth]{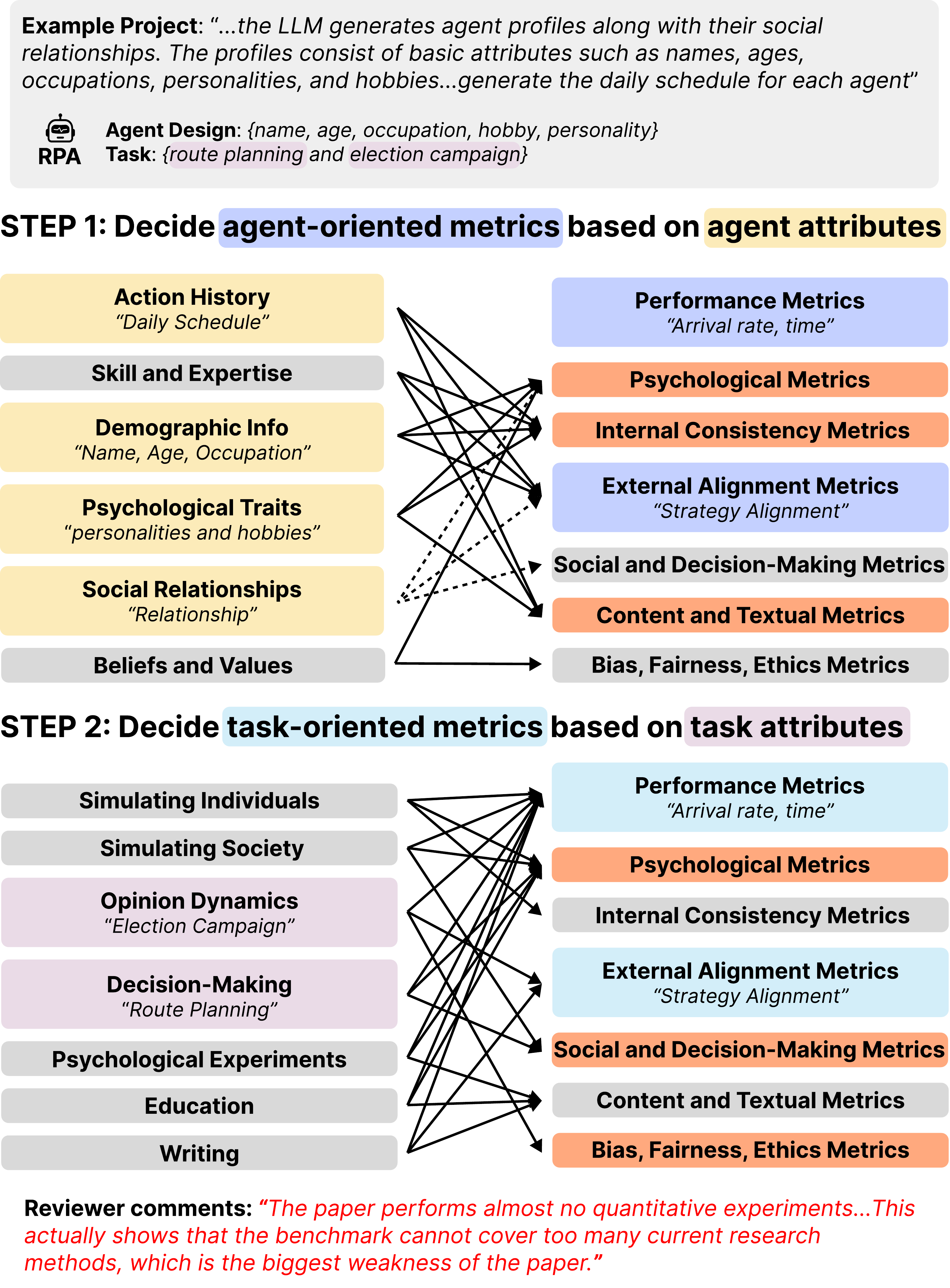}
    \caption{Case study of a flawed example in Section \ref{sec_bad_example}. Given agent attributes (yellow) and task attributes (pink). The original authors' selection of evaluation metrics (purple and blue). The missing metrics that are recommended by our proposed guideline (orange) align with the reviewer's criticism in red text.}
    \label{fig:bad example}
\end{figure}

Fig.~\ref{fig:bad example} visualized how the authors in the flawed example selected their evaluation metrics how further evaluation metrics could be uncovered through our proposed guideline.

\section{Metrics Glossary}
\label{sec: metrics glossary}

We present two glossary tables for referencing the source of agent-oriented metrics (Tab.~\ref{tab: long_table_agent_metrics_src}) and task-oriented metrics (Tab.~\ref{tab: long_table_task_metrics_src}).

\onecolumn

\begin{small}
\begin{center}
\begin{longtable}{@{}p{0.12\textwidth}p{0.22\textwidth}p{0.3\textwidth}p{0.06\textwidth}p{0.2\textwidth}@{}}
\caption{Agent-oriented evaluation metrics glossary.} \label{tab: long_table_agent_metrics_src} \\

\toprule 
\textbf{Attribute} & \textbf{Category} & \textbf{Agent-oriented Metrics} & \textbf{Approach}  & \textbf{Source} \\ \midrule 
\endfirsthead

\\
\midrule 
\textbf{Attribute} & \textbf{Category} & \textbf{Agent-oriented Metrics} & \textbf{Approach}  & \textbf{Source} \\ \midrule
\endhead

\hline \multicolumn{5}{c}{{Continued on next page}} \\ \hline
\endfoot

\hline \bottomrule
\endlastfoot

Belief \& Value        & Bias, fairness, ethics metrics      & Exaggeration (normalized average cosine similarity)                                                                                         & Automatic         & \cite{cheng-etal-2023-compost}                                                                                                                                \\
Belief \& Value        & Bias, fairness, ethics metrics      & Individuation (classification accuracy)                                                                                                     & Automatic         & \cite{cheng-etal-2023-compost}                                                                                                                                \\
Belief \& Value        & Bias, fairness, ethics metrics      & Bias (performance disparity, prevalence, magnitude, variation, attitude shift)                                                              & Automatic         & \cite{gupta2024bias}\\

Belief \& Value        & Bias, fairness, ethics metrics      & Bias (performance disparity, prevalence, magnitude, variation, attitude shift)                                                              & Automatic         & \cite{taubenfeld-etal-2024-systematic}                                                                                                                              \\
Demographic Information & Bias, fairness, ethics metrics      & Exaggeration (normalized average cosine similarity)                                                                                         & Automatic         & \cite{cheng-etal-2023-compost}                                                                                                                               \\
\end{longtable}
\end{center}
\end{small}
\twocolumn

\onecolumn

\begin{small}
\begin{center}
\begin{longtable}{@{}p{0.1\textwidth}p{0.24\textwidth}p{0.3\textwidth}p{0.06\textwidth}p{0.2\textwidth}@{}}
\caption[Task-oriented evaluation metrics glossary]{Task-oriented evaluation metrics glossary.} \label{tab: long_table_task_metrics_src} \\

\toprule
\textbf{Task} & \textbf{Category} & \textbf{Task-oriented Metrics} & \textbf{Approach}  & \textbf{Source} \\ \midrule 
\endfirsthead

\\
\midrule 
\textbf{Task} & \textbf{Category} & \textbf{Task-oriented Metrics} & \textbf{Approach}  & \textbf{Source} \\ \midrule
\endhead

\hline \multicolumn{5}{c}{{Continued on next page}} \\ \hline
\endfoot

\hline \bottomrule
\endlastfoot

Decision Making          & Social and economic metrics         & Negotiation (Concession Rate, Negotiation Success Rate, Average Negotiation Round)                                                                                                                          & Automatic & \cite{huang2024personality}                                                                                                                                                                                                                                                                                                                                                                                                 \\
Decision Making          & Social and economic metrics         & Societal Satisfaction (average per-capita living area size, average waiting time, social welfare)                                                                                                           & Automatic & \cite{ji2024srap}                                                                                                                                                                                                                                                                                                                                                                                   \\
\end{longtable}
\end{center}
\end{small}

\clearpage
\twocolumn

\end{document}